\documentclass[12pt]{article}
\usepackage{epsfig}

\begin{document}

\begin{center}

\vspace*{1.0cm}

{\Large \bf{First observation of $\alpha$ decay of $^{190}$Pt to the
first excited level ($E_{exc}=137.2$ keV) of $^{186}$Os}}

\vskip 1.0cm

{\bf
P.~Belli$^{a}$,
R.~Bernabei$^{a,b,}$\footnote{Corresponding author.
   {\it E-mail address:} rita.bernabei@roma2.infn.it (R.~Bernabei).},
F.~Cappella$^{c,d}$,
R.~Cerulli$^{e}$,
F.A.~Danevich$^{f}$,
A.~Incicchitti$^{c}$,
M.~Laubenstein$^{e}$,
S.S.~Nagorny$^{f}$,
S.~Nisi$^{e}$,
O.G.~Polischuk$^{f}$,
V.I.~Tretyak$^{f}$
}

\vskip 0.3cm

$^{a}${\it INFN, Sezione di Roma ``Tor Vergata'', I-00133 Rome, Italy}

$^{b}${\it Dipartimento di Fisica, Universit$\grave{a}$ di Roma ``Tor Vergata'' I-00133 Rome, Italy}

$^{c}${\it INFN, Sezione di Roma ``La Sapienza'', I-00185 Rome, Italy}

$^{d}${\it Dipartimento di Fisica, Universit$\grave{a}$ di Roma La ``Sapienza'', I-00185 Rome, Italy}

$^{e}${\it INFN, Laboratori Nazionali del Gran Sasso, 67010 Assergi (AQ), Italy}

$^{f}${\it Institute for Nuclear Research, MSP 03680 Kyiv, Ukraine}

\end{center}

\vskip 0.5cm

\begin{abstract}

The alpha decays of naturally occurring platinum isotopes, which are accompanied by the emission
of $\gamma$ quanta, have been  searched for deep underground (3600 m w.e.)
in the Gran Sasso National Laboratories of the INFN (Italy). A sample of Pt with mass of 42.5 g
and a natural isotopic composition has been measured with a low background HP Ge detector (468 cm$^3$)
during 1815 h. The alpha decay of $^{190}$Pt to the first excited level of 
$^{186}$Os ($J^\pi = 2^+$, $E_{exc}=137.2$ keV) has been observed for the first time,
with the half-life determined as: 
$T_{1/2} = 2.6_{-0.3}^{+0.4}$(stat.)$\pm0.6$(syst.)$\times10^{14}$ yr. The
$T_{1/2}$ limits for the $\alpha$ decays of other Pt isotopes have been determined at level of 
$T_{1/2} \simeq 10^{16}-10^{20}$ yr. These limits have been set for the first time or they are 
better than those known from earlier experiments.

\end{abstract}

\vskip 0.4cm


\vskip 0.4cm


\section{Introduction}

The phenomenon of the $\alpha$ decay is already known since more than 100 years \cite{Rut99}
but the interest in this process is still great; in fact, it has even been increased during
the last decade, both from the theoretical and the experimental sides. 
Almost 30 theoretical papers can be found in the literature published only in 2009 -- 2010.
Many of them were devoted to new semi-empirical formulae for half-lives (see f.e. \cite{Den09})
which successfully describe the accumulated to-date experimental $T_{1/2}^\alpha$ values
and allow one to understand in a more clear way the perspectives of new investigations,
in particlular, in studies of exotic very short-living isotopes close to the proton
drip line \cite{Gro08} or decays of super-heavy elements \cite{Hof09}.
In the investigations of long-living rare nuclear decays, the field of interest of the authors of this 
article,
the improvements in the experimental techniques led to the enhancement of the sensitivity
and to the discovery of new $\alpha$ decays which were never observed previously because of the
extra long half-lives of the decaying nuclides:
$T_{1/2}^\alpha = 1.9\times10^{19}$ yr was found for $^{209}$Bi which was considered before as 
the heaviest stable nuclide \cite{Mar03};
half-lives in the range of 
$T_{1/2}^\alpha = (1.0-1.8)\times10^{18}$ yr were measured  for the $^{180}$W in \cite{Dan03,Coz04,Zde05,Bel10};
the $\alpha$ decay of $^{151}$Eu was also recently observed with 
$T_{1/2}^\alpha = 5.0\times10^{18}$ yr \cite{Bel07}.  The
$^{209}$Bi and the $^{180}$W are keepers of two current world records: 
the $^{209}$Bi for the longest $\alpha$ half-life measured so far, and the 
$^{180}$W for the lowest observed specific $\alpha$ activity of only 2.3 disintegrations per year
per gram of W of natural composition which is much lower than that for $^{209}$Bi
(105 disintegrations per year per gram of Bi),
due to low natural abundance of the $^{180}$W ($\delta=0.12\%$ 
comparing to $\delta=100\%$ for $^{209}$Bi \cite{Boh05}).  

All the six naturally occurring isotopes of platinum are potentially unstable in relation to $\alpha$
decay (see Table 1). However, only for one of them, the $^{190}$Pt (with the biggest energy release of
$Q_\alpha = 3251(6)$ keV), this process was experimentally observed to-date, with the first successful
measurement in 1921 \cite{Hof21}. In that and in subsequent works the half-life of the $^{190}$Pt was determined in
the range of $(3.2-10)\times10^{11}$ yr (see review \cite{Tav06} and refs. therein); the currently recommended
half-life value is:  $T^\alpha_{1/2}=(6.5\pm0.3)\times10^{11}$ yr \cite{Bag03}. 
In all the previous works, the $^{190}$Pt $\alpha$ decay was
observed only to the ground state (g.s.) of $^{186}$Os. However, the first excited level of the daughter
nuclide $^{186}$Os ($J^\pi = 2^+$) has a quite low energy: $E_{exc} = 137.2$ keV \cite{ToI98}, 
and the energy available to the
$\alpha$ particle in the decay to this level: $Q_\alpha^* = 3114(6)$ keV, is not much lower than that 
in the g.s. to g.s. transition. Our theoretical estimates of the corresponding half-life (see details in Section
4) gave values in the range of $T^\alpha_{1/2}=10^{13}-10^{14}$ yr. 
This allowed us to hope to discover the $^{190}$Pt $\to$ $^{186}$Os($2^+_1$) decay 
through the observation of the 137.2 keV $\gamma$ quantum emitted in the deexcitation of
the $^{186}$Os$^*$ nucleus with a well shielded low background HP Ge detector even using a Pt sample
with natural isotopic composition  
with very low percentage of $^{190}$Pt ($\delta = 0.014\%$).

\begin{table}[htb]
\caption{Natural abundances ($\delta$) of Pt isotopes and energy releases ($Q_\alpha$) expected in the $\alpha$
decays of Pt to Os. $N_i$ is number of nuclei of the specific Pt isotope per 1 g of natural platinum.}
\begin{center}
\begin{tabular}{|llll|}
\hline
Parent      & $\delta$, \% & $Q_\alpha$, keV & $N_i/1$ g \\
isotope     & \cite{Boh05} & \cite{Aud03}    &           \\
\hline
$^{190}$Pt  & 0.014(1)     & 3251(6)         & $4.32\times10^{17}$ \\
$^{192}$Pt  & 0.782(7)     & 2418.6(2.2)     & $2.41\times10^{19}$ \\
$^{194}$Pt  & 32.767(99)   & 1518.3(1.6)     & $1.01\times10^{21}$ \\
$^{195}$Pt  & 33.832(10)   & 1172.0(1.6)     & $1.04\times10^{21}$ \\
$^{196}$Pt  & 25.242(41)   & 808.1(2.6)      & $7.79\times10^{20}$ \\
$^{198}$Pt  & 7.163(55)    & 100(4)          & $2.21\times10^{20}$ \\
\hline
\end{tabular}
\end{center}
\label{tb:QI}
\end{table}

The present paper describes the first observation of the $\alpha$ decay $^{190}$Pt 
$\to$ $^{186}$Os($2^+_1$) based on the 
measurements with a low background HP Ge detector operated in the underground conditions of
the Gran Sasso National Laboratories (LNGS) of the INFN (Italy). In addition to the $^{190}$Pt decay, the $\gamma$ quanta can
be emitted in the $\alpha$ decays of other Pt isotopes: 
(1) when the excited levels of the daughter Os nuclei are populated; 
(2) when the daughter Os nuclei are unstable and further decay with emission
of some $\gamma$ rays, as in the case of $^{195}$Pt and $^{198}$Pt. We also determine 
$T_{1/2}$ limits for such decays (to our knowledge, for the first time).

\section{Measurements}

The measurements were performed deep underground at LNGS; the average overburden is about 3600 meters of water equivalent. Two
platinum crucibles, exploited in the LNGS chemical laboratory, were used as the Pt sample with
the total mass of 42.53 g. 
The bigger Pt cup (see Fig. 1a) has the maximal diameter of near 6.4 cm, height 2.8 cm and 
thickness around 0.21 mm. Approximate sizes of the smaller Pt crucible are 3.0 cm (maximal diameter),
3.2 cm (height) and 0.25 mm (thickness).

\nopagebreak
\begin{figure}[htbp]
\begin{center}
\mbox{\epsfig{figure=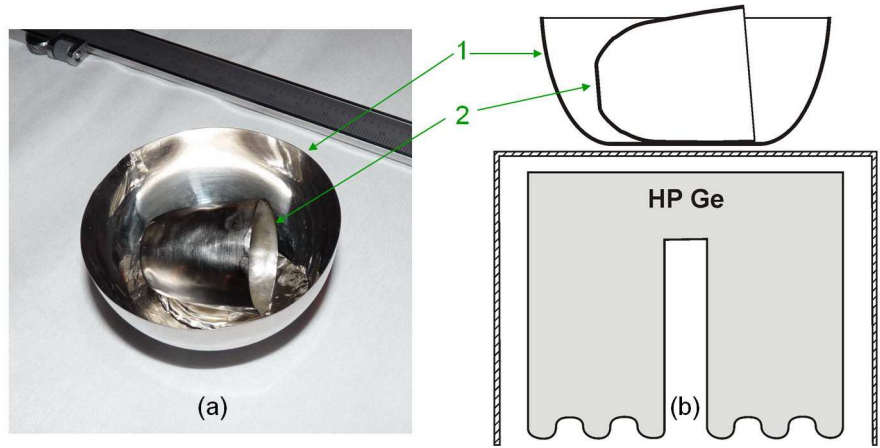,height=6.0cm}}
\caption{(a) Photo of the Pt sample and (b) simplified scheme of the measurements
with the HP Ge detector. The bigger Pt cup and smaller Pt crucible 
are labeled as 1 and 2, respectively.}
\end{center}
\end{figure}

The data with the Pt sample were collected with HP Ge detector (GeCris, 468 cm$^3$; 
see Fig. 1b for simplified scheme of the measurements) over 1815.4 h.
The Pt sample in thin polyethylene bag was placed directly on the lid of the detector's cryostat.
The background spectrum of the detector was measured during 1045.6 h. 
The energy resolution of the detector is FWHM = 2.0 keV for the 1332 keV $\gamma$ line of $^{60}$Co.
To reduce the external background, the detector was shielded by layers of low-radioactive copper
($\simeq10$ cm) and lead ($\simeq20$ cm). The set-up has been continuously flushed by high purity
boil-off nitrogen to avoid the presence of residual environmental radon.
Part of the spectrum accumulated with the Pt sample in comparison with the background in
the energy range of 100 -- 700 keV is shown in Fig. 2.

\nopagebreak
\begin{figure}[htb]
\begin{center}
\mbox{\epsfig{figure=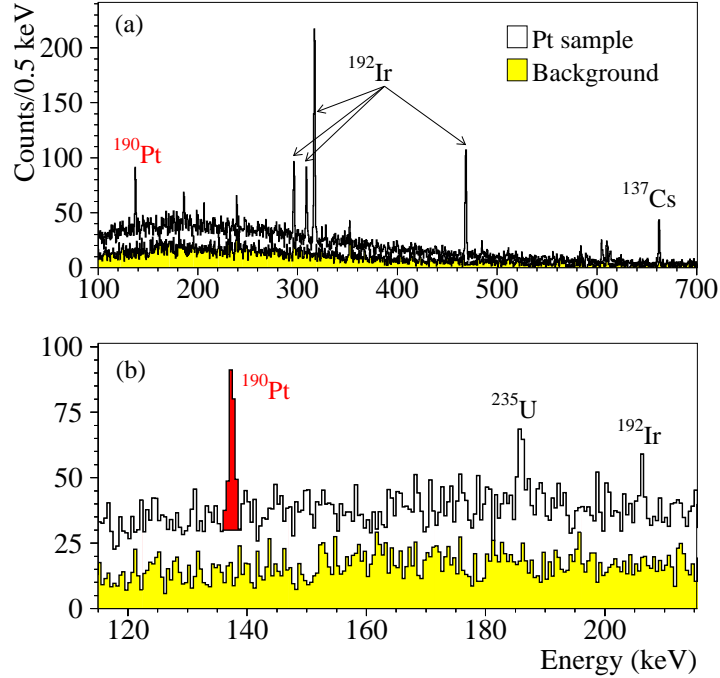,height=9.0cm}}
\caption{(Color online) Energy spectrum of the Pt sample with mass of 42.5 g
measured during 1815 h in the 100 -- 700 keV energy interval (a),
and in more detail around the 137 keV region (b).
The background spectrum (measured during 1046 h but normalized here to 1815 h) 
is also shown (filled histogram). 
Peak at 137 keV after $\alpha$ decay $^{190}$Pt $\to$ $^{186}$Os$(2^+_1)$
is clearly visible in the Pt spectrum being absent in the background.}
\end{center}
\end{figure}

The comparison of the Pt and of the background spectra shows that the platinum sample practically 
is not contaminated by ``usual'' radioactive contaminants: U/Th series, $^{40}$K, $^{60}$Co, 
(for which mostly only limits are determined at the level of $\simeq10$ mBq/kg; see details in 
accompanied paper on the search for double $\beta$ processes in Pt \cite{Bel11}). 
However, the presence of the radioactive $^{192}$Ir
is evident; the corresponding activity is equal to ($49\pm3$) mBq/kg. The $^{192}$Ir could appear in platinum as
cosmogenic activation of Pt by cosmic rays at the Earth surface. In fact, because iridium usually
accompanies platinum in nature, $^{192}$Ir can be created in result of neutron capture by
$^{191}$Ir which is one of two naturally occurring iridium isotopes ($\delta=37.3\%$, cross section for 
thermal neutrons is $\sigma=954$ b \cite{ToI98}). However, while half-life of the ground
state of $^{192}$Ir is $T_{1/2}=73.8$ d \cite{ToI98}, and the exponential decrease in time 
of the $^{192}$Ir activity should be observed during our 75.6 d measurements, in fact, 
the rate of the corresponding peaks was
consistent with constant. Thus, the $^{192}$Ir activity should be ascribed not to the
ground state, but to the isomeric $^{192m}$Ir level with $E_{exc}=168.1$ keV and $T_{1/2}=241$ yr 
\cite{Bag98}. This isomeric state decays to the ground state of $^{192}$Ir emitting 155.1 keV 
and 13.0 keV $\gamma$ rays. The transitions, however, are strongly converted to electrons (coefficients of 
conversion are equal $\alpha_{13}=57000$ and $\alpha_{155}=1026$ \cite{Bag98}).
This explains the absence of the 155.1 keV peak in our data (Fig. 2).

\section{Results}

\subsection{$\alpha$ decay of $^{190}$Pt}

In the spectrum collected with the Pt sample, the peak at energy ($137.1\pm0.1$) keV is
observed (see Fig. 2) while it is absent in the background spectrum. 
The fit of the Pt spectrum in the energy region (120 -- 170) keV with a Gaussian
peak and linear background assumption gives a net area of ($132\pm17$) counts, inconsistent with
zero at about $8\sigma$. Variations of the energy interval for the fit result in 
changes of the area inside the quoted uncertainty.

The 137.1 keV peak can be explained with the $\alpha$ decay of $^{190}$Pt to the first
excited level of $^{186}$Os whose excitation energy is ($137.159\pm0.008$) keV \cite{Bag03}.
If populated, this level deexcites with the emission of a $\gamma$ quantum with energy
$E_\gamma=(137.157\pm0.008)$ keV which is in nice agreement with that of the observed 
peak ($137.1\pm0.1$) keV. The process of $\alpha$ decay $^{190}$Pt $\to$ $^{186}$Os$^*$
was never observed previously. 

Using the area of the 137 keV peak, the corresponding partial
half-life for the transition to the first excited level of $^{186}$Os can be calculated as:
\begin{equation}
T_{1/2}(^{190}\mbox{Pt} \to~ ^{186}\mbox{Os}(2^+_1, 137.2~\mbox{keV})) =
\frac{\ln 2 \cdot N_{190} \cdot \varepsilon \cdot t}{S \cdot (1+\alpha)},
\end{equation}
where $N_{190}=1.84\times10^{19}$ is the number of the $^{190}$Pt nuclei in the 42.53 g Pt sample, 
$\varepsilon$ is the efficiency to detect the full energy $\gamma$ with the HP Ge detector, 
$t=1815.4$ h is the measurement time,
$S=(132\pm17)$ counts is the area of the peak and 
$\alpha$ is the coefficient of conversion to electrons for the given nuclear transition.

The full peak efficiency at 137 keV was calculated with the EGS4 \cite{EGS4} and,
for crosscheck, also with GEANT4 \cite{GEANT4} simulation packages. 
The Pt crucibles have non-trivial shapes, and models used in each code (programmed
independently) reproduced the real shape with some approximations. The calculations
gave for the 137 keV peak values of $\varepsilon=3.4\%$ with EGS4 and 
$\varepsilon=2.6\%$ with GEANT4. For the final result, we use average between
these two values $\varepsilon=3.0\%$ including the difference into systematic uncertainties.

Taking into account the electron conversion coefficient
for the transition $\alpha=1.29$ \cite{Bag03}, the $T_{1/2}$ value is equal
$T_{1/2}(^{190}$Pt $\to$ $^{186}$Os$(2^+_1, 137.2$ keV)) = 
$2.6_{-0.3}^{+0.4}$(stat.)$\times10^{14}$ yr.

Only the statistical uncertainty in the peak area was taken here into account. The
systematic uncertainties are related with the uncertainty 
of the mass of the Pt sample (0.02\%), 
of the calculation of the measurements' live time (0.01\%),
of the knowledge on natural isotopic abundance of $^{190}$Pt (7.1\%),
with the biggest contribution from the calculation of the efficiency.
To estimate the latter uncertainty, we performed a comparison of the calculated and measured
efficiencies for a voluminous ($\oslash7.0\times1.1$ cm) water source in which, among several
radioactive isotopes, also $^{57}$Co was dissolved. 
This source was calibrated and distributed by the International Atomic Energy Agency (IAEA) 
within an open world-wide proficiency test. The radioactive $^{57}$Co, in particular,
emits $\gamma$ quanta with the energy of 136.5 keV, very close to the peak 
under investigation. Disagreement between the experimental and calculated efficiencies
was 6\% (0\%) with EGS4 (GEANT4). 
Taking into account the previous experience of measurements with the used GeCris HP Ge detector
and the more complicated geometry of the Pt sample in comparison
with simple cylindrical shape of the IAEA water source, 
the systematic uncertainty in the efficiency can be conservatively estimated as 20\%.

Summing all the uncertainties in squares, we obtain the following value for the half-life of
$\alpha$ decay $^{190}$Pt $\to$ $^{186}$Os$(2^+_1)$ as the final:

\begin{equation}
T_{1/2}(^{190}\mbox{Pt} \to~ ^{186}\mbox{Os}(2^+_1, 137.2~\mbox{keV})) =
2.6_{-0.3}^{+0.4}\mbox{(stat.)}\pm0.6\mbox{(syst.)}\times10^{14} ~\mbox{yr}.
\end{equation}

It should be noted that $^{192}$Ir, present in the Pt sample, also emits
$\gamma$ rays with energy of 136.3 keV, however, with very low yield of $I=0.183\%$ \cite{ToI98}. 
Taking into account that the area of the most intensive ($I=82.80\%$) peak of $^{192}$Ir at
316.5 keV is ($619\pm32$) counts (and considering also different efficiencies of
2.3\% at 137 keV\footnote{The efficiency for the 137 keV peak of $^{192}$Ir is lower than that
for the single 137 keV $\gamma$ quantum because of summing effects for $^{192}$Ir
$\gamma$ rays emitted in cascade.}
and 5.5\% at 316 keV), contribution of $^{192}$Ir to the 137 keV peak is 
0.6 counts; this does not change the $T_{1/2}$ value presented in Eq. (2).

Gamma rays with energies close to 137 keV are emitted in some other nuclear 
processes (see f.e. \cite{WWW}), 
and this could give an alternative explanation of the peak observed in the Pt
experimental spectrum; however, usually additional $\gamma$ rays are
also emitted in such decays, and their absence allows us to exclude these effects. 
For example, $^{181}$Hf ($T_{1/2}=42.4$ d \cite{ToI98}), 
which could be created in result of cosmogenic activation of Pt,
emits 136.3 keV ($I=5.85\%$) and 136.9 keV ($I=0.86\%$) $\gamma$ quanta but
also 133.0 keV $\gamma$ rays should be emitted with much higher yield of $I=43.3\%$
which, however, are absent in the experimental data (see Fig. 2).
Other example could be the 136.6 keV ($I=0.012\%$) $\gamma$ rays from $^{235}$U  
but the peak at 143.8 keV ($I=10.96\%$) is absent. The contributions from the nuclides of the U/Th
natural radioactive chains: $^{214}$Pb ($E_\gamma=137.5$ keV, $I<0.006\%$) and
$^{228}$Ac ($E_\gamma=137.9$ keV, $I=0.024\%$) are calculated as negligible
using the data on U/Th pollution of the Pt sample.

$^{181}$W ($T_{1/2}=121.2$ d), other possible cosmogenic contamination of Pt, cannot be estimated in
the above-described way because its 136.3 keV $\gamma$ rays are the most intensive
with yield of $I=0.0311\%$. We calculated the $^{181}$W induced activity in Pt with the COSMO code 
\cite{Mar92}; the result was 1.5 decays per day per kg of $^{nat}$Pt. 
Taking into account the time of measurements (75.6 d), the small mass of our Pt sample (42.5 g) and the 
low yield of these $\gamma$ rays, the contribution from $^{181}$W to the 137 keV peak
is estimated also as negligible ($5\times10^{-5}$ counts).

In conclusion, analysing other possible sources of the 137 keV line, we did not
find real alternative which could mimic $\alpha$ decay 
$^{190}$Pt $\to$ $^{186}$Os$(2^+_1)$. 
The old scheme of the $^{190}$Pt $\alpha$ decay \cite{Bag03,ToI98} and the updated scheme
which follows from our observation of the $^{190}$Pt $\to$ $^{186}$Os$(2^+_1)$ transition
is presented in Fig. 3.

\nopagebreak
\begin{figure}[htbp]
\begin{center}
\mbox{\epsfig{figure=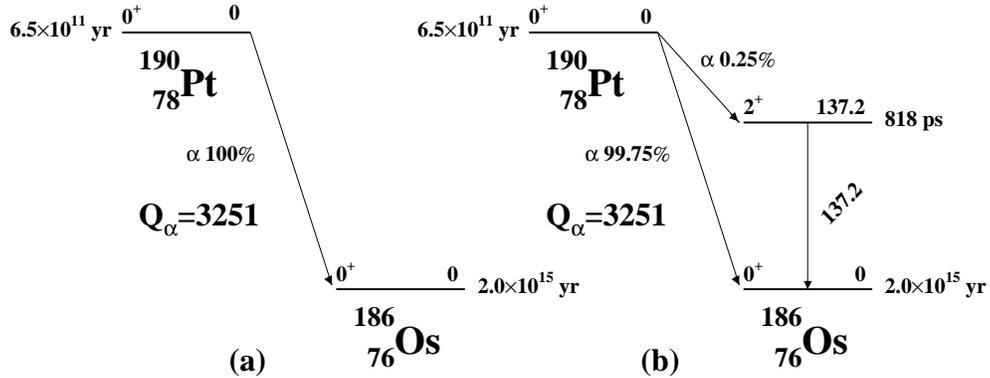,height=5.0cm}}
\caption{Old (a) and new (b) schemes of $\alpha$ decay of $^{190}$Pt.
The energies of the levels and the deexcitation $\gamma$ quantum are given 
in keV \cite{Bag03,ToI98}.}.
\end{center}
\end{figure}

In the process of $^{190}$Pt $\alpha$ decay other excited levels of $^{186}$Os could be 
populated too. Because the probability of the $\alpha$ decay exponentially decreases with the decrease
of the energy, we consider here possible transition only to the next excited level of 
$^{186}$Os ($J^\pi=4^+$) with $E_{exc}=434.1$ keV. If this level is populated, two $\gamma$ quanta are
emitted in its deexcitation with energies of $E_{\gamma1}=296.9$ keV and $E_{\gamma2}=137.2$ keV.
Both peaks are present in our data, and while we explain the 137 keV peak by the 
$^{190}$Pt $\to$ $^{186}$Os$(2^+_1)$ $\alpha$ decay, the 296 keV peak is related with 
$^{192}$Ir. 
We can estimate the contribution to the 296 keV peak from $^{192}$Ir using its nearby peak at 
316.5 keV. Taking into account that the area of the 316 keV peak is ($619\pm32$) counts, the
yields $I_{316}=82.80\%$ and $I_{296}=28.67\%$ \cite{ToI98}, we could expect ($193\pm10$) 
counts from $^{192}$Ir. At the same time, in the real Pt spectrum the area of the 296 keV peak
is $S=(200\pm23)$ counts. The difference between these two values ($7\pm25$) counts is consistent 
with 0 and, in accordance with the Feldman-Cousins procedure \cite{Fel98}, results in  the
limit: $S<48$ counts at 90\% C.L. 
Deriving (here and in the following) the $T_{1/2}$ limits for the Pt $\alpha$ decays, 
to be more conservative, we will use for the efficiencies the values obtained 
with the code giving systematically lower efficiency.
Substituting in Eq. (1) the value of the
electron conversion coefficient $\alpha=0.095$ \cite{Bag03} and the
efficiency $\varepsilon=7.1\%$ for 296 keV $\gamma$ quanta,
we obtain the following half-life limit:
$$T_{1/2}(^{190}\mbox{Pt} \to~ ^{186}\mbox{Os}(4^+_1, 434.1~\mbox{keV})) >
3.6\times10^{15} ~\mbox{yr at 90\% C.L.}$$

\subsection{$T_{1/2}$ limits on $\alpha$ decays of other Pt isotopes}

The data collected with the Pt sample also allow us to search for $\alpha$ decays of 
other Pt isotopes related with the emission of $\gamma$ quanta. We will consider 
here only transitions to the lowest excited levels of the daughter Os isotopes as the
most probable. In the case of $^{195}$Pt and $^{198}$Pt, the daughter Os nuclei are unstable,
and the search for $\gamma$ quanta related with their decays gives the possibility to
look for the g.s. to g.s. transitions of Pt to Os. 
In general, we do not see any of such decays,
and only limits on the corresponding half-lives are determined.

{\bf $^{192}$Pt.} The best half-life limit on $\alpha$ decay of $^{192}$Pt was obtained in
\cite{Kau66} searching for its specific activity: $T_{1/2}>6.0\times10^{16}$ yr. 
This value could be considered as valid not only for the g.s. to g.s. transition, but 
also for the decay to the lowest excited levels of $^{188}$Os.
 
The population of the first excited level of $^{188}$Os ($J^\pi=2^+$, $E_{exc}=155.0$ keV)
will lead to the emission of $\gamma$ quantum and to a peak at energy of 155.0 keV.
The latter, however, is absent in the Pt spectrum (see Fig. 2), and we can give 
only a limit on its area.
The value of $\lim S$ was calculated fitting the experimental Pt spectrum by the sum of a linear
function (representing the near linear background) and a gaussian (representing the 
expected effect) with center at 155.0 keV and
proper FWHM close to that of the nearby 137 keV peak. 
This procedure gave the value $S=(-8\pm17)$ counts for the area of the 155 keV peak that results 
in the limit \cite{Fel98} $\lim S = 21$ counts at 90\%~C.L.
Using the number of $^{192}$Pt nuclei: $N_{192}=1.02\times10^{21}$, the electron conversion
coefficient $\alpha=0.82$ \cite{Sin02} and the calculated efficiency $\varepsilon=3.5\%$, 
one gets with Eq. (1):
$$T_{1/2}(^{192}\mbox{Pt} \to~ ^{188}\mbox{Os}(2^+_1, 155.0~\mbox{keV})) >
1.3\times10^{17} ~\mbox{yr at 90\% C.L.}$$

The alpha decay of $^{192}$Pt to the next excited level of $^{188}$Os ($J^\pi=4^+$, 
$E_{exc}=477.9$ keV) will lead to the emission of two deexcitation $\gamma$ quanta
with energies of 322.9 keV and 155.0 keV. The peak at energy of 322.9 keV is absent
in the data: its area found in a procedure similar to the one described above gave 
$S=(12\pm16)$ counts that results in $\lim S = 38$ counts at 90\%~C.L. The electron conversion 
coefficient for the 322.9 keV transition is equal $\alpha=0.074$ \cite{Sin02},
and efficiency is $\varepsilon=7.2\%$. With these values we obtain:
$$T_{1/2}(^{192}\mbox{Pt} \to~ ^{188}\mbox{Os}(4^+_1, 477.9~\mbox{keV})) >
2.6\times10^{17} ~\mbox{yr at 90\% C.L.}$$

{\bf $^{194}$Pt.} We give below limits for possible $^{194}$Pt $\alpha$ decays to the
two lowest excited levels of $^{190}$Os: 
$J^\pi=2^+$, $E_{exc}=186.7$ keV and $J^\pi=4^+$, $E_{exc}=547.9$ keV. 
Their population results in the emission of one $\gamma$ quantum with $E_\gamma=186.7$ keV,
or two $\gamma$ rays with $E_{\gamma1}=186.7$ keV, $E_{\gamma2}=361.1$ keV, correspondingly.

There is a peak at energy ($185.6\pm0.1$) keV in the experimental Pt spectrum with area
$S=(79\pm16)$ counts; most 
probably it belongs to $^{235}$U which emits $\gamma$ rays of 185.7 keV with yield of 57.20\%.
It is difficult to estimate the contribution of $^{235}$U to the 186 keV peak using 
other $\gamma$ lines in the $^{235}$U chain because they have lower yields. 
Thus, to derive limit for $\alpha$ decay $^{194}$Pt $\to$ $^{190}$Os$(2^+_1)$,
we very conservatively ascribe all the peak area to the process of $\alpha$ decay;
taking also into account the uncertainty in the $S$ value, we obtain $\lim S = 105$ counts at 90\% C.L.
Substituting in Eq. (1) the number of $^{194}$Pt nuclei: $N_{194}=4.30\times10^{22}$, the  
efficiency $\varepsilon=4.9\%$ and the conversion coefficient $\alpha=0.425$ \cite{Sin03},
one finally obtains:
$$T_{1/2}(^{194}\mbox{Pt} \to~ ^{190}\mbox{Os}(2^+_1, 186.7~\mbox{keV})) >
2.0\times10^{18} ~\mbox{yr at 90\% C.L.}$$

The peak at the energy of 361.1 keV, expected for $^{194}$Pt $\to$ $^{190}$Os$(4^+_1)$ decay,
is absent, and using the values $S=(-9.2\pm8.3)$ counts  (thus $\lim S =6.2$ counts), 
$\varepsilon=7.2\%$ and $\alpha=0.054$ \cite{Sin03}, one gets:
$$T_{1/2}(^{194}\mbox{Pt} \to~ ^{190}\mbox{Os}(4^+_1, 547.9~\mbox{keV})) >
6.8\times10^{19} ~\mbox{yr at 90\% C.L.}$$

{\bf $^{195}$Pt.} The daughter nucleus $^{191}$Os is unstable ($T_{1/2}=15.4$ d \cite{ToI98})
and $\beta^-$ decays to $^{191}$Ir. This allows us to set the $T_{1/2}$ limit for $^{195}$Pt $\alpha$
decay not only to some specific excited level of $^{191}$Os but to all the $^{191}$Os states
(including g.s.).
In the $^{191}$Ir decay, a $\gamma$ quantum with energy of 129.4 keV
and yield of $I=29.0\%$ is emitted \cite{Van07}. Peak at this energy is absent in the Pt 
spectrum (see Fig. 2), its area $S=(-15.7\pm10.9)$ counts gives $\lim S = 6.4$ counts.
Taking into account the number of the $^{195}$Pt nuclei: $N_{195}=4.42\times10^{22}$,
and the efficiency $\varepsilon=2.2\%$, we obtain:
$$T_{1/2}(^{195}\mbox{Pt} \to~ ^{191}\mbox{Os, all states})) >
6.3\times10^{18} ~\mbox{yr at 90\% C.L.}$$

{\bf $^{196}$Pt.} We give here the $T_{1/2}$ limit only for the decay to the lowest excited level
of $^{192}$Os ($J^\pi=2^+$, $E_{exc}=205.8$ keV). It should be noted that in the experimental
Pt spectrum we have a peak at energy of ($205.9\pm0.1$) keV and area of $S=(36\pm12)$ counts.
However, it is related with the contamination of the Pt sample by $^{192}$Ir which emits
$\gamma$ rays of 205.8 keV with yield $I_{206}=3.30\%$ \cite{ToI98}.
We can estimate contribution of $^{192}$Ir to the 206 keV peak on the basis of another peak of 
$^{192}$Ir at 316.5 keV ($I_{316}=82.80\%$). Taking into account that $S_{316}=(619\pm32)$ counts
and the efficiencies $\varepsilon_{316}=5.5\%$, $\varepsilon_{206}=4.3\%$, we can expect
($19\pm1$) counts in the 206 keV peak from $^{192}$Ir. 
Contribution from $^{235}$U, which emits $\gamma$ rays of 205.3 keV with $I=5.01\%$, can be
estimated from its 185.7 keV peak ($I=57.20\%$) as: ($8\pm2$) counts. 
The difference in areas between the line observed in the experimental spectrum and the contributions
from $^{192}$Ir and $^{235}$U is equal to ($9\pm12$) counts giving no evidence for the effect.
With the number of $^ {196}$Pt nuclei: $N_{196}=3.31\times10^{22}$, $\lim S = 29$ counts, the efficiency
$\varepsilon=5.5\%$ and the electron conversion coefficient $\alpha=0.305$ \cite{Bag98},
we obtain:
$$T_{1/2}(^{196}\mbox{Pt} \to~ ^{192}\mbox{Os}(2^+_1, 205.8~\mbox{keV})) >
6.9\times10^{18} ~\mbox{yr at 90\% C.L.}$$

{\bf $^{198}$Pt.} The daughter $^{194}$Os nucleus is unstable ($T_{1/2}=6.0$ yr \cite{Sin06})
and decays to $^{194}$Ir which is unstable too. In the decay chain
$^{194}$Os (6.0 yr, $Q_\beta=96.9$ keV) $\to$ 
$^{194}$Ir (19.28 h, $Q_\beta=2233.8$ keV) $\to$ 
$^{194}$Pt (stable) 
the most prominent $\gamma$ line is at 328.5 keV with yield of $I=13.1\%$.
Peak at this energy is absent in the Pt spectrum: $S=(12\pm9)$ counts, $\lim S = 27$ counts. 
Using the number of $^ {198}$Pt nuclei: $N_{198}=9.40\times10^{21}$, and the efficiency
$\varepsilon=7.2\%$, one gets:
$$T_{1/2}(^{198}\mbox{Pt} \to~ ^{194}\mbox{Os, all states})) >
4.7\times10^{17} ~\mbox{yr at 90\% C.L.}$$

As in the case of $^{195}$Pt, this limit is valid for the $\alpha$ decay of $^{198}$Pt
to all the states of $^{194}$Os including the g.s. to g.s. transition. A summary of all the
obtained results is given in Table 2.

\begin{table}[htb]
\caption{Summary of $T_{1/2}$ results. Limits correspond to 90\% C.L.
Theoretical half-lives are calculated in accordance with receipts of \cite{Buc91} and \cite{Poe83}
taking into account additional hindrance factors.
For $^{195}$Pt and $^{198}$Pt, the experimental $T_{1/2}$ limits are valid for the $\alpha$ decays to
all the states of the daughter Os nuclei while the theoretical estimations are given for the g.s. to 
g.s. transitions.}
\begin{center}
\begin{tabular}{|lllll|}
\hline
Alpha                       & Level of daughter  & \multicolumn{2}{l}{Experimental $T_{1/2}$ (yr)}              & Theoretical \\
transition                  & nucleus            & Previous                         & This work                 & estimation  \\
\hline
~ & ~ & ~ & ~ & ~ \\
$^{190}$Pt $\to$ $^{186}$Os & $2^+$, 137.2 keV   & --                               & $=2.6\pm0.7\times10^{14}$ & $3.2\times10^{13}$  \cite{Buc91} \\
~                           &                    &                                  &                           & $7.0\times10^{13}$  \cite{Poe83} \\
~                           & $4^+$, 434.1 keV   & --                               & $>3.6\times10^{15}$       & $7.4\times10^{17}$  \cite{Buc91} \\
~                           &                    &                                  &                           & $2.0\times10^{18}$  \cite{Poe83} \\
$^{192}$Pt $\to$ $^{188}$Os & $2^+$, 155.0 keV   & $>6.0\times10^{16}$ \cite{Kau66} & $>1.3\times10^{17}$       & $9.1\times10^{25}$  \cite{Buc91} \\
~                           &                    &                                  &                           & $4.6\times10^{26}$  \cite{Poe83} \\
~                           & $4^+$, 477.9 keV   & $>6.0\times10^{16}$ \cite{Kau66} & $>2.6\times10^{17}$       & $2.7\times10^{33}$  \cite{Buc91} \\
~                           &                    &                                  &                           & $1.9\times10^{34}$  \cite{Poe83} \\
$^{194}$Pt $\to$ $^{190}$Os & $2^+$, 186.7 keV   & --                               & $>2.0\times10^{18}$       & $7.0\times10^{51}$  \cite{Buc91} \\
~                           &                    &                                  &                           & $1.9\times10^{53}$  \cite{Poe83} \\
~                           & $4^+$, 547.9 keV   & --                               & $>6.8\times10^{19}$       & $4.3\times10^{71}$  \cite{Buc91} \\
~                           &                    &                                  &                           & $3.5\times10^{73}$  \cite{Poe83} \\
$^{195}$Pt $\to$ $^{191}$Os & all states         & --                               & $>6.3\times10^{18}$       & $7.5\times10^{59}$  \cite{Buc91} \\
~                           &                    &                                  &                           & $3.2\times10^{69}$  \cite{Poe83} \\
$^{196}$Pt $\to$ $^{192}$Os & $2^+$, 205.8 keV   & --                               & $>6.9\times10^{18}$       & $1.7\times10^{106}$ \cite{Buc91} \\
~                           &                    &                                  &                           & $2.1\times10^{109}$ \cite{Poe83} \\
$^{198}$Pt $\to$ $^{194}$Os & all states         & --                               & $>4.7\times10^{17}$       & $1.6\times10^{348}$ \cite{Buc91} \\
~                           &                    &                                  &                           & $7.2\times10^{358}$ \cite{Poe83} \\
\hline
\end{tabular}
\end{center}
\label{tb:summary}
\end{table}

\section{Comparison with theoretical estimations}

We calculated half-life values for the $\alpha$ decays of the different Pt isotopes with
the cluster model of Ref. \cite{Buc91} and with semiempirical
formulae \cite{Poe83} based on liquid drop model and description of
$\alpha$ decay as a very asymmetric fission process. The approaches
\cite{Buc91,Poe83} were tested with a set of experimental half-lives of
almost four hundred $\alpha$ emitters and demonstrated good agreement between
calculated and experimental $T_{1/2}$ values, mainly inside the factor of $2-3$.
We also successfully used these works to predict $T_{1/2}$ values in our
searches for $\alpha$ decays of $^{180}$W \cite{Dan03} and $^{151}$Eu \cite{Bel07}
obtaining adequate agreement between the first experimentally measured and
calculated results.
As other example, the calculated $T_{1/2}$ values for the g.s. to g.s. $\alpha$
decay of $^{190}$Pt are equal: 
$5.0\times10^{11}$ yr in accordance with \cite{Buc91} and 
$10.3\times10^{11}$ yr with \cite{Poe83}, 
while the experimental values were measured in the range of $(3.2-10)\times10^{11}$ yr \cite{Tav06}
(recommended value is $(6.5\pm0.3)\times10^{11}$ yr \cite{Bag03}).

The ground state of all the Pt isotopes has spin and parity $J^\pi=0^+$, except of
$^{195}$Pt with $J^\pi=1/2^-$. We study here the transitions to the daughter levels with
$J^\pi=2^+$ and $4^+$ ($9/2^-$ for $^{195}$Pt $\to$ $^{191}$Os g.s.).
Because of the difference between the spins of the initial and of the final nuclear states,
the emitted $\alpha$ particle will have non-zero angular momentum: values of $L=2$
or $L=4$ are the most preferable among the allowed ones by the selection rules for the 
transitions with $\Delta J^{\Delta \pi}=2^+$ or $4^+$, respectively.
Thus, in the $T_{1/2}$ estimation, we take into account the
hindrance factors of HF = 1.9 for $L=2$ and HF = 9.0 for $L=4$, 
calculated in accordance with \cite{Hey99}.
The results of the calculations are given in Table 2.
For the $\alpha$ decay $^{190}$Pt $\to$ $^{186}$Os($2^+$, 137.2 keV) observed in this work, the
semiempirical approach of \cite{Poe83} gives a better agreement with the measured $T_{1/2}$
value.

From the theoretical results, it is clear that the observation of other Pt $\alpha$ decays
presented in Table 2 is out of reach of current experimental possibilities, except
probably for decay $^{190}$Pt $\to$ $^{186}$Os($4^+$, 434.1 keV). The latter
will need 3 orders of magnitude higher sensitivity which, however, could be possible
by using enriched $^{190}$Pt instead of natural Pt with $\delta(^{190}$Pt) = 0.014\%.

\section{Conclusions}

The alpha decays of naturally occurring platinum isotopes, which are accompanied by the emission
of $\gamma$ quanta, were searched for with the help of a low background HP Ge detector
in the Gran Sasso National Laboratories of the INFN.
The alpha decay of $^{190}$Pt to the first excited level of $^{186}$Os ($J^\pi = 2^+$, $E_{exc}=137.2$ keV)
was detected for the first time by the observation of the $\gamma$ line at energy ($137.1\pm0.1$) keV
at $8\sigma$ significance level while it is absent in the background spectrum of the detector.
We did not find alternative processes which could mimic this effect.
The half-life is determined as: 
$T_{1/2}(^{190}$Pt $\to$ $^{186}$Os$(2^+_1, 137.2$ keV)) =
[$2.6_{-0.3}^{+0.4}$(stat.)$\pm0.6$(syst.)]$\times10^{14}$ yr.
This value is in relevant agreement with theoretical calculations
based on the liquid drop model and the description of the $\alpha$ decay as a very asymmetric
fission process \cite{Poe83}.

In addition, $T_{1/2}$ limits for $\alpha$ decays with population of the lowest excited levels 
of Os nuclei 
(for $^{192}$Pt, $^{194}$Pt, $^{196}$Pt) and for transition to the ground states of Os nuclei
(for $^{195}$Pt, $^{198}$Pt) were determined at level of $T_{1/2} \simeq 10^{16}-10^{20}$ yr.
These limits were set for the first time or they are better than those known from earlier
experiments.

The alpha decay of $^{190}$Pt to the second excited level of $^{186}$Os ($J^\pi = 4^+$, $E_{exc}=434.1$ keV)
could be detected in higher sensitivity experiment with platinum enriched in $^{190}$Pt
to $\simeq10\%$. However, because its natural abundance is extremely low ($\delta = 0.014\%$), this is a very
expensive task with the current technologies. 

\section{Acknowledgements}

The authors would like to thank V.Yu. Denisov for interesting discussions 
and anonymous referee for useful suggestions.
The group from the Institute for Nuclear Research (Kyiv, Ukraine) was supported in part 
through the Project ``Kosmomikrofizyka-2'' (Astroparticle Physics) of the National
Academy of Sciences of Ukraine.

\end{document}